\documentclass[11pt]{article}

\usepackage{psfig}
\usepackage[dvips]{epsfig}
\usepackage{latexsym}     
\usepackage{amssymb} 
\usepackage{graphicx}
\usepackage[figuresright]{rotating}

\newcommand{\be}{\begin{equation}} 
\newcommand{\ee}{\end{equation}}   
\newcommand{\bea}{\begin{eqnarray}}
\newcommand{\eea}{\end{eqnarray}}
\newcommand{\bean}{\begin{eqnarray*}}
\newcommand{\eean}{\end{eqnarray*}}  
\newcommand{\bef}{\begin{figure}}  
\newcommand{\eef}{\end{figure}}  
\newcommand{\bet}{\begin{table}}
\newcommand{\eet}{\end{table}}

\begin{document}

\title{\bf{On the continuum limit of gauge-fixed compact $U(1)$ lattice
gauge theory}}
\author
{Subhasish Basak$^a$\thanks{email: sbasak@physics.umd.edu},
Asit K De$^b$\thanks{email: de@theory.saha.ernet.in} and
Tilak Sinha$^b$\thanks{email: tilak@theory.saha.ernet.in} \\
\\
$^a$Department of Physics, University of Maryland,\\
College Park, MD 20742, USA\\
$^b$Theory Group, Saha Institute of Nuclear Physics,\\
1/AF, Salt Lake, Calcutta 700064, India}
\date{}
\maketitle

\begin{abstract}
We investigate the continuum limit of a compact formulation of the lattice
U(1) gauge theory in 4 dimensions using a nonperturbative gauge-fixed
regularization. We find clear evidence of a continuous phase transition in 
the pure gauge theory for all values of the gauge coupling (with gauge
symmetry restored). When probed with quenched staggered fermions with U(1)
charge, the theory clearly has a chiral transition for large gauge couplings.
We identify the only possible region in the parameter space where a
continuum limit with nonperturbative physics may appear.
\end{abstract}

\section*{Introduction}
Although quantum field theories were first formulated on the lattice
regulator to investigate the nonperturbative properties of 
Quantum Chromodynamics, the lattice regulator can be useful in general to
study nonperturbative behavior of any field theory, in particular the
theories involving nonasymptotically-free couplings. In this paper, we have
looked at Quantum Electrodynamics (QED) which on the lattice can be
formulated in terms of compact group-valued gauge fields in the usual Wilson
approach, or in terms of noncompact gauge fields as in the
continuum. The noncompact formulation does not allow any nonperturbative
behavior in the pure gauge sector (it does show nonperturbative behavior
only in the presence of fermions). However, the compact formulation allows
for self-interaction of gauge fields on the lattice regulator and hence it
is interesting to study its phase structure and possible continuum limits.
Obviously for it to be a viable regularization, the lattice theory must have
a weak coupling continuum limit that would produce free photons in the pure
gauge sector. Once it has the expected weak coupling continuum limit, one is
interested in finding a continuum limit with possible nonperturbative
properties.

It is well known that compact formulation of U(1) gauge theory (with or
without fermions) on the lattice has at least two different phases: a weak 
coupling phase (with usual QED-like properties on the regulator) called the 
Coulomb phase and a strong coupling confining phase which resembles QCD 
(again on the lattice regulator) in many ways - 
existence of gauge balls, confinement, nonzero chiral 
condensate, appearance of Goldstone bosons etc. To remove the regulator,
there ought to be a continuous phase transition in the intermediate
region. Numerical studies have found the existence of such a phase transition 
\cite{azco0,jiri0}, but the order of the phase transition is generally accepted 
as first order.
   
A continuum limit from the confinement phase sustaining
the properties of the confinement phase
would imply a hitherto unknown continuum abelian gauge theory 
which is likely to correspond to  a non-trivial fixed point.      

As far as continuum limits from the Coulomb phase is concerned,
one would have to imagine that the critical manifold (obtained by
appropriately expanding the parameter space) is not too far away  
from the weak coupling region
since perturbative results on lattice match excellently with our 
familiar continuum QED.
However, the lack of a continuum limit for compact lattice QED in the weak
coupling region means that the issue
of triviality (existence of Landau poles etc.) cannot really be answererd in
a genuine (nonperturbative) way.  

With the addition of a nonminimal plaquette term in the gauge action 
\cite{jiri}, there is new evidence that the Coulomb-confinement transition
in the pure gauge theory is first order \cite{flux}.  
Only with inclusion of fermions (with a four
fermion interaction \cite{azco1,kogut}), there is indication of a 
continuous phase transition.

In this paper we make an exploratory study of possible continuum limits of 
a compact lattice formulation of pure
U(1) gauge theory using a different regularization in 4 space-time dimensions. 
We also probe the pure gauge theory 
by quenched staggered fermions which have U(1)-charge.
This regularization of lattice U(1) theory
was originally devised to `tame' the `rough gauge' problem of lattice chiral 
gauge theories \cite{reduced}. Because of the gauge-invariant measure and 
the lack of gauge-invariance of the lattice chiral
gauge theories,
the longitudinal gauge degrees of freedom ({\em lgdof}) couple  
nonperturbatively to the physical degrees of freedom. To decouple the
{\em lgdof} which are radially frozen scalar fields, a nonperturbative gauge
fixing scheme (corresponding to a local renormalizable covariant gauge
fixing in the naive continuum limit) for the compact U(1) gauge fields was
proposed \cite{golter}. A key feature of this gauge fixing scheme is that
the gauge fixing term is not the exact square of the expression used in the
gauge fixing condition and as a result not BRST-invariant (as required by
Neuberger's theorem \cite{neu} for compact gauge fixing). It has, in 
addition, appropriate irrelevant terms to make the perturbative vacuum 
$U_{\mu x}=1$ unique. Because the gauge fixing term obviously breaks gauge
invariance, one needs to add counterterms to restore manifest gauge
symmetry.

The parameter space of this regularization of compact U(1)
theory now includes the gauge coupling, the coefficient of the gauge fixing
term and the coefficients of the counterterms. In this extended parameter
space it has been shown \cite{bock1} that for a weak gauge coupling
($g=0.6$) there exists 
a continuous phase transition at which the U(1) gauge
symmetry is restored and the continuum theory of free photons emerge.
In this study, determination of the phase diagram and the critical region 
as extensively as possible is very important because
we are dealing with a gauge-noninvariant theory in general and it is
necessary to have freedom along the critical manifold so that irrelevant
parameters can be appropriately tuned to restore gauge invariance
(protecting the theory from a violation of unitarity).

Scanning a wide range of all the three parameters, in the pure gauge theory
we have found phase transitions between a phase with broken rotational
symmetry (FMD phase) and one with rotational symmetry (FM phase). The FM-FMD
transition is the place where the gauge symmetry gets restored. To recover
Lorentz invariance in the continuum, the FM-FMD transition needs to be
approached from the FM side. 

Probing the pure gauge theory with
quenched U(1)-charged fermions, we have also found evidence for a
chiral transition for large gauge couplings although in this exploratory study
we could assess
its approximate location in a limited region of the parameter space.
We have looked for chiral condensates only near the FM-FMD transition,
because this is where the gauge symmetry is
restored. 

From our numerical evidence we expect that the chiral phase transition 
intersects the FM-FMD phase
transition at the tricritical line where the order of the FM-FMD
transition changes. 
The region of the FM-FMD transition where the chiral condensate is nonzero
seems to be first order. The intersection region of the chiral
and the FM-FMD transitions thus seems to be the only candidate
where a continuum limit with nonperturbative properties may be achieved.

A preliminary account of our work can be found in \cite{pre}.

\section*{The Regularization}

The action for the compact gauge-fixed U(1) theory 
\cite{golter}, where the ghosts are free and decoupled, is:
\be
S[U] = S_g[U] + S_{gf}[U] + S_{ct}[U]. \label{action}
\ee

$S_g$ is the usual Wilson plaquette action,
\be
S_G = \frac{1}{g^2} \sum_{x\,\mu <\nu} \left( 1 -
{\rm Re}\, U_{\mu\nu x} \right) \label{plact}
\ee
where $g$ is the gauge coupling and $U_{\mu x}$ is the group valued U(1)
gauge field. 

$S_{gf}$ is the BRST-noninvariant compact gauge fixing term, 
\be
S_{gf} = \tilde{\kappa} \left( \sum_{xyz} \Box(U)_{xy}
\Box (U)_{yz} - \sum_x B_x^2 \right) \label{gfact}
\ee
where $\tilde{\kappa}$ is the coefficient of the gauge fixing term, 
$\Box(U)$ is the covariant lattice Laplacian and
\be
B_x  = \sum_\mu \left( \frac{{\cal A}_{\mu, x-\mu} +
{\cal A}_{\mu x}}{2}\right)^2,
\ee
where ${\cal A}_{\mu x} {\rm = Im}\, U_{\mu x}$. As mentioned in the
Introduction, $S_{gf}$ is not
just a naive transcription of the continuum covariant gauge fixing  
term, it has in addition appropriate irrelevant terms. This makes the action 
have an unique absolute minimum at $U_{\mu x}= 1$,
validating weak coupling perturbation theory around $g=0$ or
$\tilde{\kappa}=\infty$ and in the naive continuum limit reduces
to $1/2\xi \int d^4x (\partial_\mu A_\mu)^2$ with $\xi =
1/(2{\tilde{\kappa}}g^2)$. 

Validity of weak  
coupling perturbation theory together with perturbative
renormalizability helps to determine the form of the counter
terms to be present in $S_{ct}$. It turns out that the most
important gauge counterterm is the dimension-two counterterm, namely the 
gauge field mass counterterm
given by,
\be
S_{ct} = - \kappa \sum_{\mu x} \left( U_{\mu x} +
U_{\mu x}^\dagger \right). 
\ee
In the pure bosonic theory there are possible marginal counterterms including
derivatives. However, in the investigation of the gauge-fixed theory as
given, the dimension-two counterterm has been mostly considered, because it
alone could lead to a continuous phase transition that recovers the gauge
symmetry. It was argued that the marginal counterterms would not possibly
create new universality classes for the continuum theory corresponding to
large $\tilde{\kappa}$ (for a discussion on other counterterms, please see
\cite{golter,bock1}). 

Our philosophy here has been an usual one, i.e. to take a lattice 
theory given by
(\ref{action}) having the expected  weak coupling results, and then
try and find out the strong coupling properties of the same theory.

Information about the phase diagram of the model can be obtained in
the constant field approximation, first by expanding the link field
$U_{\mu x}= \exp igA_{\mu x}$ around $U_{\mu x}=1$  and then requiring
the gauge potential $A_\mu$ to be constant (thus all the terms
containing derivatives of $A_\mu$ vanish). Since the WCPT is defined
around $U_{\mu x}=1$, the classical potential $V_{cl}$ is the leading
order approximation of the effective potential and is given by,
\bea
V_{cl} &=& \kappa \left[ g^2 \sum_\mu A_\mu^2 +
\cdots \right] + \nonumber\\
&& \frac{g^4}{2\xi} \left[ \left( \sum_\mu A_\mu^2 \right)
\left( \sum_\mu A_\mu^4 \right) + \cdots \right]. \label{vcl}
\eea
where the dots represent terms of higher order in $g^2$. Since
the perturbation theory is defined around $g=0$ or $\tilde{\kappa}
= \infty$, the classical potential is expected to be reliable at
least for the region of large $\tilde{\kappa}$.

From (\ref{vcl}) it follows that for $\kappa > 0$, the gauge boson
mass is nonzero and $V_{cl}$ has a minimum at $A_\mu = 0$. The region
$\kappa>0$ therefore is a phase with broken gauge symmetry -- the FM
phase.

For $\kappa < 0$, the minimum of $V_{cl}$ shifts to a nonzero value:
\[
A_\mu = \pm \left( \frac{\xi \vert \kappa \vert}{3 g^2}
\right)^{\frac{1}{4}} \hspace{0.8cm} {\rm for~all}\;\; \mu
\]
implying an unusual phase with broken rotational symmetry in addition
to the broken gauge symmetry -- we call it the directional ferromagnetic
phase (FMD) \cite{golter}.

For $\kappa=0 \equiv\kappa_c$ the minimum of $V_{cl}$ is at $A_\mu=0$
and at the same time the gauge boson mass vanishes, thus gauge symmetry
is restored which signals a continuous phase transition or criticality.

For large enough $\tilde{\kappa}$ the presence of the $\sim A_\mu^6$
term in the gauge fixing action (\ref{gfact}) produces a continuous
phase transition FM$\searrow$FMD at which the gauge boson mass scales
to zero and hence the gauge symmetry is restored. The continuum limit
now can be taken by approaching FM-FMD transition from within FM phase.

\section*{Numerical Simulation and Results}
\bef
\vspace{-1.5cm}
\centerline{\psfig{figure=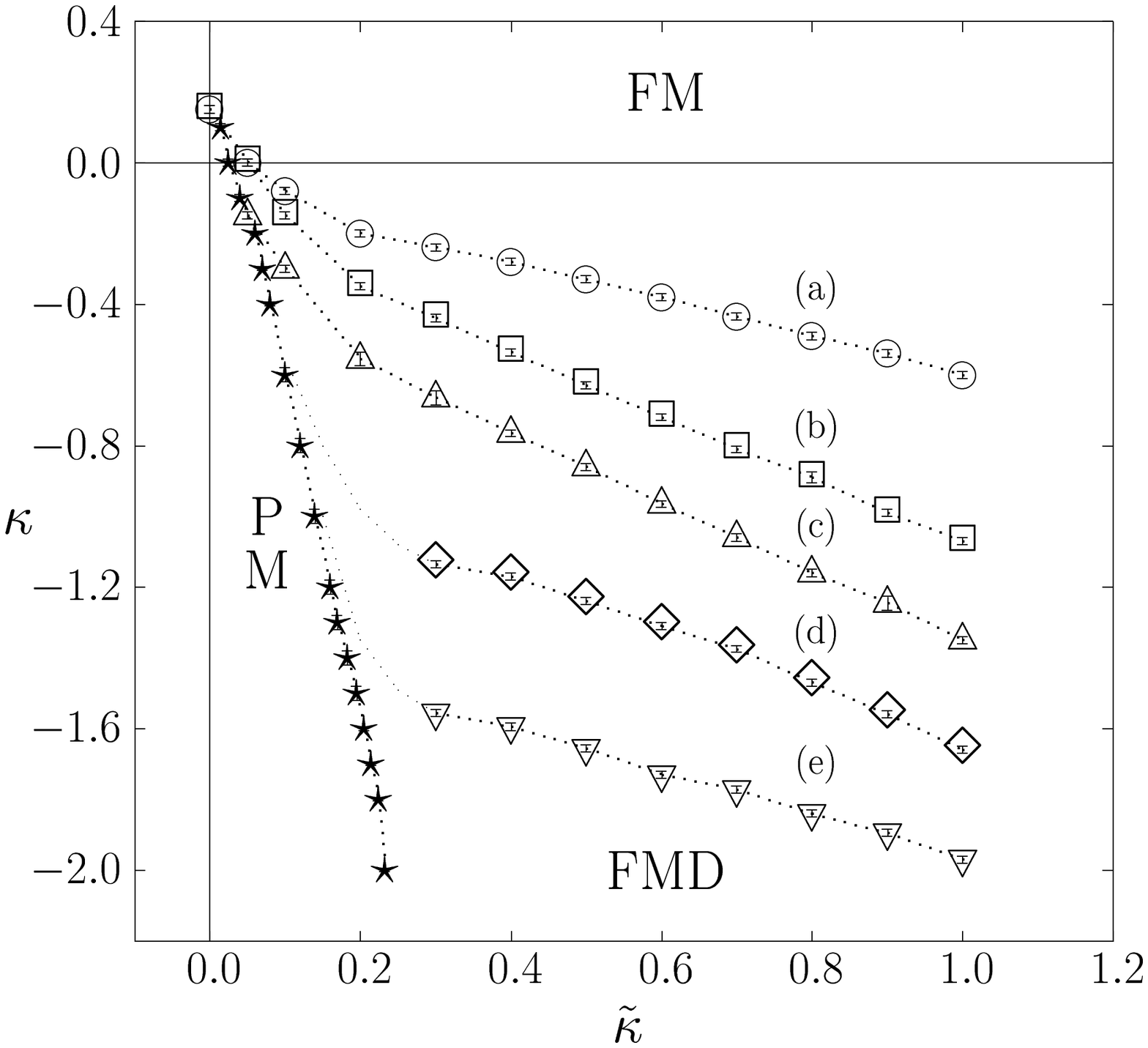,width=12.0cm,height=14cm}}
\vspace{-4.5cm}
\caption{Phase diagram of pure $U(1)$ gauge fixed
theory on $10^4$ lattice. 
FM-FMD transition for different gauge couplings:
(a) $g=0.8\, (\bigcirc)$, (b) $g=1.0\, (\Box)$, (c) $g=1.1\,
(\triangle)$, (d) $g=1.2\, (\Diamond)$, (e) $g=1.3\,
(\bigtriangledown)$. The line joining the $\star$-symbols
separates the PM phase from the FM and the FMD phases. The PM
transition line does not depend too much on the gauge coupling and
an average line is drawn here for clarity. } \label{phaseg}
\eef

To obtain the phase diagram of the gauge-fixed pure $U(1)$ theory,
given by the action (\ref{action}), in $(\kappa,\,\tilde{\kappa})$
-plane for fixed values of the gauge coupling $g$, we defined the following
observables (for a $L^4$-lattice):
\bea
E_P &=& \frac{1}{6L^4} \left\langle \sum_{x, \mu<\nu} {\rm Re}\,
U_{\mu \nu x} \right\rangle \label{obsep} \\
E_\kappa &=& \frac{1}{4L^4} \left\langle \sum_{x, \mu} {\rm Re}\,
U_{\mu x} \right\rangle \label{obsek} \\
V &=& \left\langle \sqrt{\frac{1}{4} \sum_\mu \left(\frac{1}{L^4}
\sum_x {\rm Im}\, U_{\mu x} \right)^2} \;\right\rangle. \label{obsv}
\eea

$E_P$ and $E_\kappa$ are not order parameters but they signal
phase transitions by sharp changes. We expect $E_\kappa \neq 0$
in the broken symmetric phases FM and FMD and $E_\kappa \sim 0$ 
in the symmetric (PM) phase. Besides, $E_\kappa$ is expected to be
continuous at a continuous phase transition (infinite slope in 
the infinite volume limit) and show a discrete jump at a first order    
transition \cite{bock1,bock2}. The true order parameter is $V$ which allows us to     
distinguish the FMD phase (where $V \neq 0$) from the other phases 
where $V \sim 0$.

The Monte Carlo simulations were done with a 4-hit Metropolis
algorithm on a variety of lattice sizes from $4^4$ to $16^4$,
although investigations were mostly done on $10^4$ lattices. The phase 
diagram was explored in $(\kappa, \tilde{\kappa})$-plane at gauge couplings
$g\,=\,0.6,\,0.8,\,1.0,\,1.1,\,1.2,\,1.3$ and 1.4 over a range
of 0.30 to -2.30 for $\kappa$ and 0.00 to 1.00 for $\tilde{\kappa}$.
The autocorrelation length for all observables was less than 
10 for $10^4$ lattices and each
expectation value was calculated from about a thousand independent
configurations.

Figure \ref{phaseg} collectively shows the phase diagram
in $(\kappa,\,\tilde{\kappa})$-plane for the different gauge
couplings. The diagram looks qualitatively the same for
all gauge couplings. For zero or small values of ${\tilde{\kappa}}$, there is
a FM-PM transition. The FM-FMD transition which ensures recovery of the
gauge symmetry is obtained 
at larger finite values of the coupling ${\tilde{\kappa}}$. As seen from the
figure, the FM-FMD transition can be approached by tuning just one
parameter, namely, $\kappa$ for given values of $g$ and ${\tilde{\kappa}}$.     

The phase diagram and behavior of the theory at 
$g=0.6$ was investigated before \cite{bock1}. We have reconfirmed the results
of \cite{bock1} and for reasons of clarity fig.1 does not include the data
for $g=0.6$. At weaker gauge couplings $g\,=\,0.6,\,0.8,
\,1.0$ the FM-FMD transition (the dotted lines roughly parallel 
to the $\tilde{\kappa}$-axis) appears to be continuous. On the other hand,
for stronger gauge couplings $g\,=\,1.1,\,1.2,\,1.3$ and also $g=1.4$ 
(although not shown in fig.1) this transition is first order for smaller
values of $\tilde{\kappa}$ and is continuous for larger values of
$\tilde{\kappa}$. The critical value of $\tilde{\kappa}$ at which the order
changes shifts to larger values with increasing gauge coupling. In this
exploratory study we have not determined the precise value of the above
mentioned critical $\tilde{\kappa}$ for the whole range of gauge couplings
investigated.
 
The order of the FM-FMD phase transition
is inferred from the change of $E_\kappa$ with
$\kappa$. A $E_\kappa$ versus $\kappa$ 
plot is shown here only for a stronger gauge coupling, discussed next.

Figure \ref{ordr} which depicts the nature of change of $E_\kappa$
versus $\kappa$ across the FM-FMD transition at $g=1.3$, 
${\tilde{\kappa}}=0.4$ for (a) $10^4$ and (b) $16^4$ lattices, shows a 
discrete jump,
implying a first order transition. Although the figure is presented
only for $g=1.3$, our observation is that at ${\tilde{\kappa}}\sim 0.4$ and 
$g\gtrsim 1.1$ the FM-FMD transition seems to be first order, whereas for 
$g<1.1$ and any ${\tilde{\kappa}}$ (large enough to accomodate a FM-FMD
transition) this transition is continuous. 

At large gauge couplings ($g\gtrsim 1.1$), when we increase the value 
of ${\tilde{\kappa}}$,
however, the discrete jump of $E_\kappa$, as $\kappa$ changes across the 
FM-FMD transition, disappears. This is clearly shown in fig.\ref{ordr1}
for $g=1.3$ and ${\tilde{\kappa}}=0.8$ for (a) $10^4$ and (b) $16^4$
lattices. This actually would indicate a
revival of a continuous FM-FMD transition at these couplings. From our 
experience it is reasonable to expect that for a larger gauge
coupling, the transition would still remain continuous if $\tilde{\kappa}$
is made large enough.

\bef[h]
\vspace{-1.5cm}
\parbox{8cm}{\psfig{figure=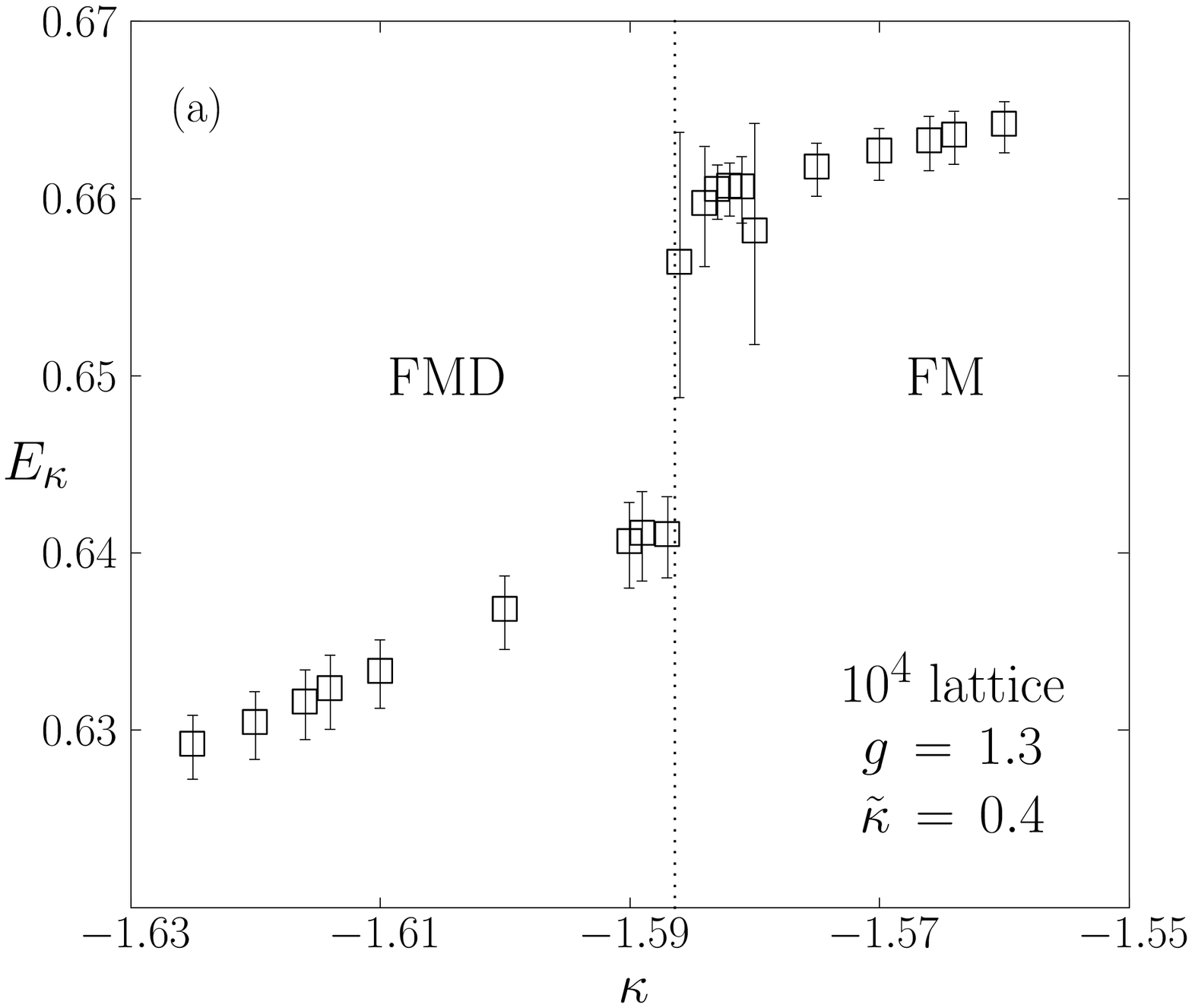,width=8.6cm,height=10.6cm}} \ \
\parbox{8cm}{\psfig{figure=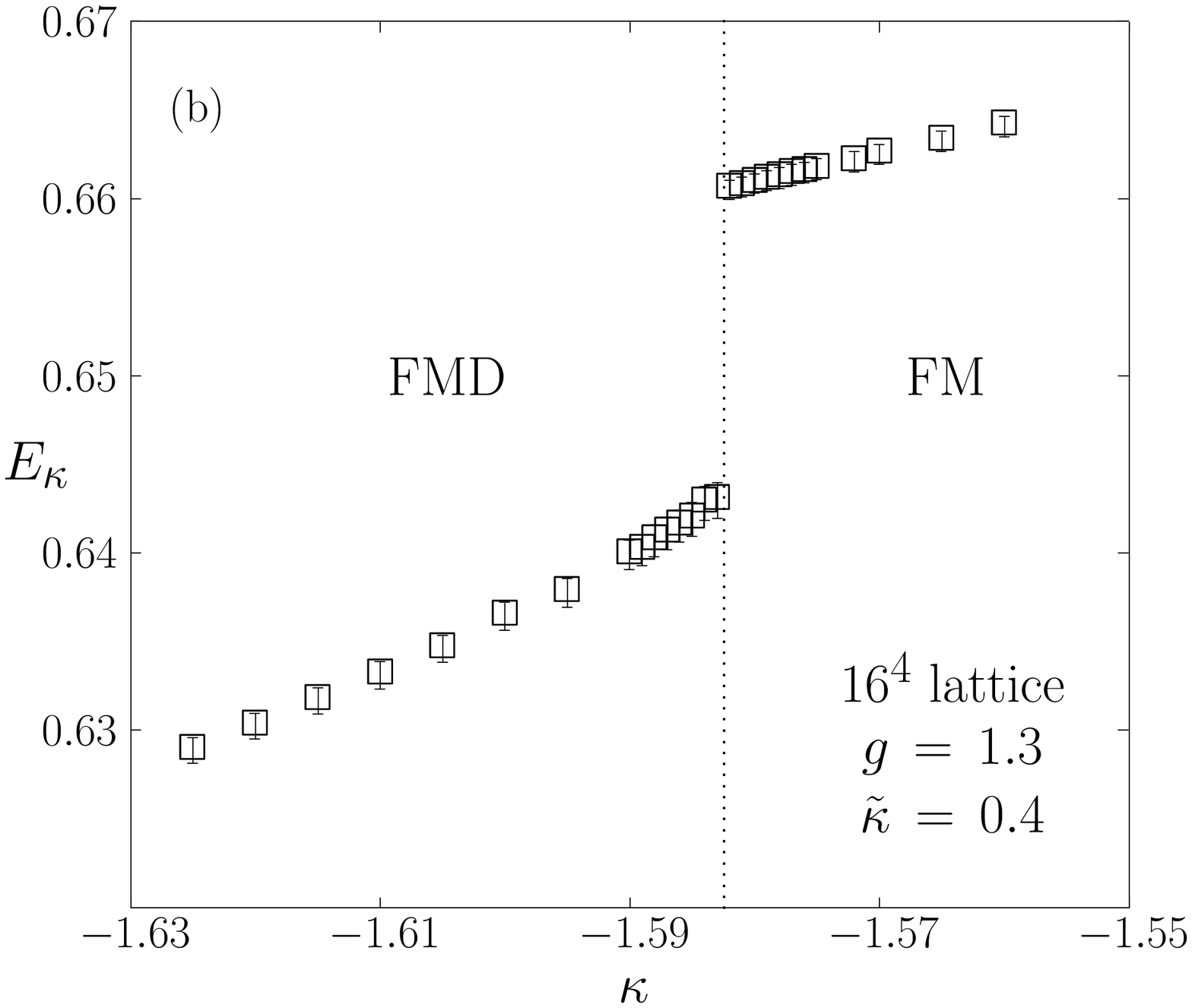,width=8.6cm,height=10.6cm}}
\vspace{-2.9cm}
\caption{Discontinuity in $E_\kappa$ (for large gauge
coupling) across FM-FMD transition (the dotted line) for (a) $10^4$ and (b)
$16^4$ lattices}
\label{ordr}
\eef

\bef[h]
\vspace{-1.5cm}
\parbox{8cm}{\psfig{figure=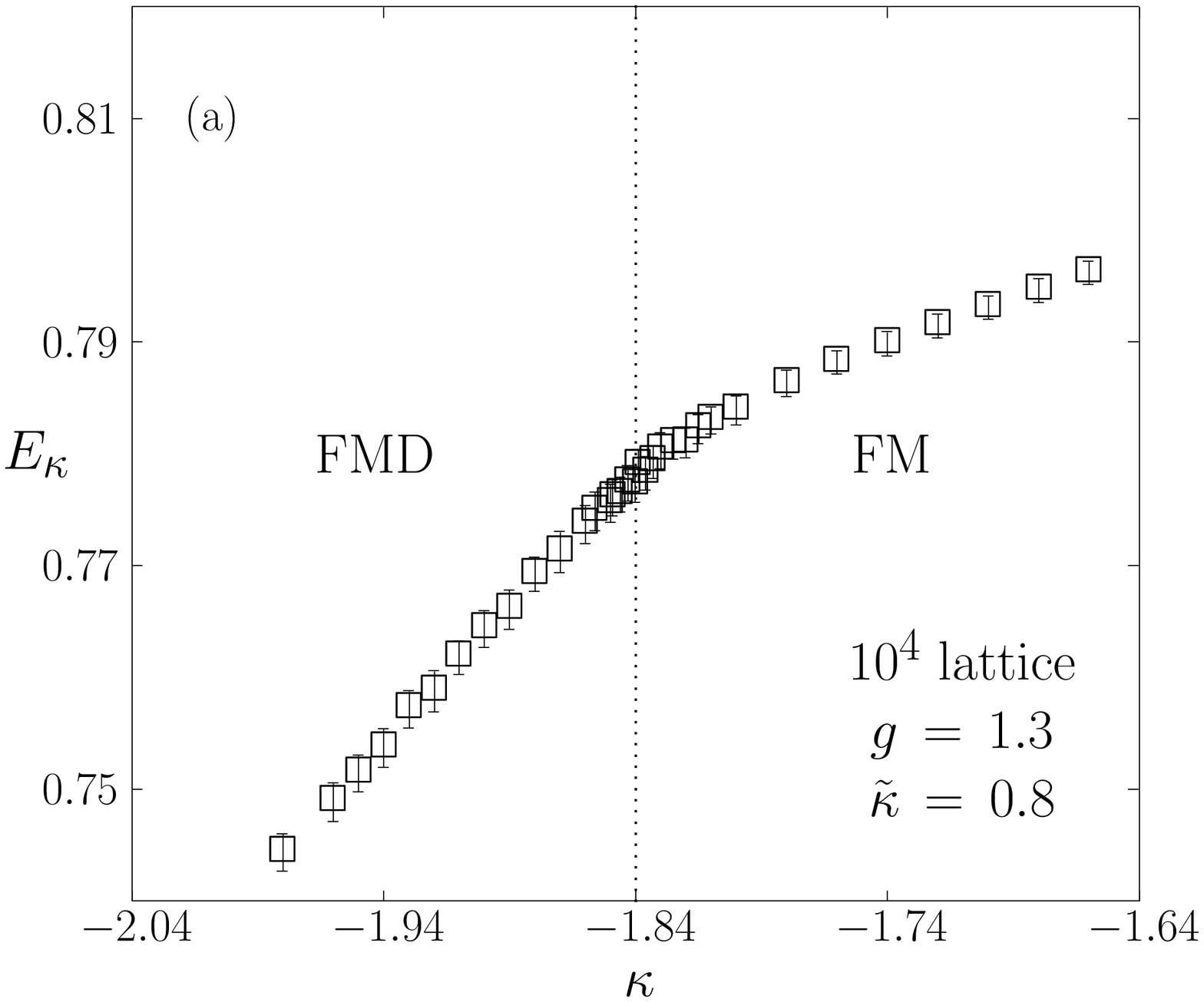,width=8.6cm,height=10.6cm}} \ \
\parbox{8cm}{\psfig{figure=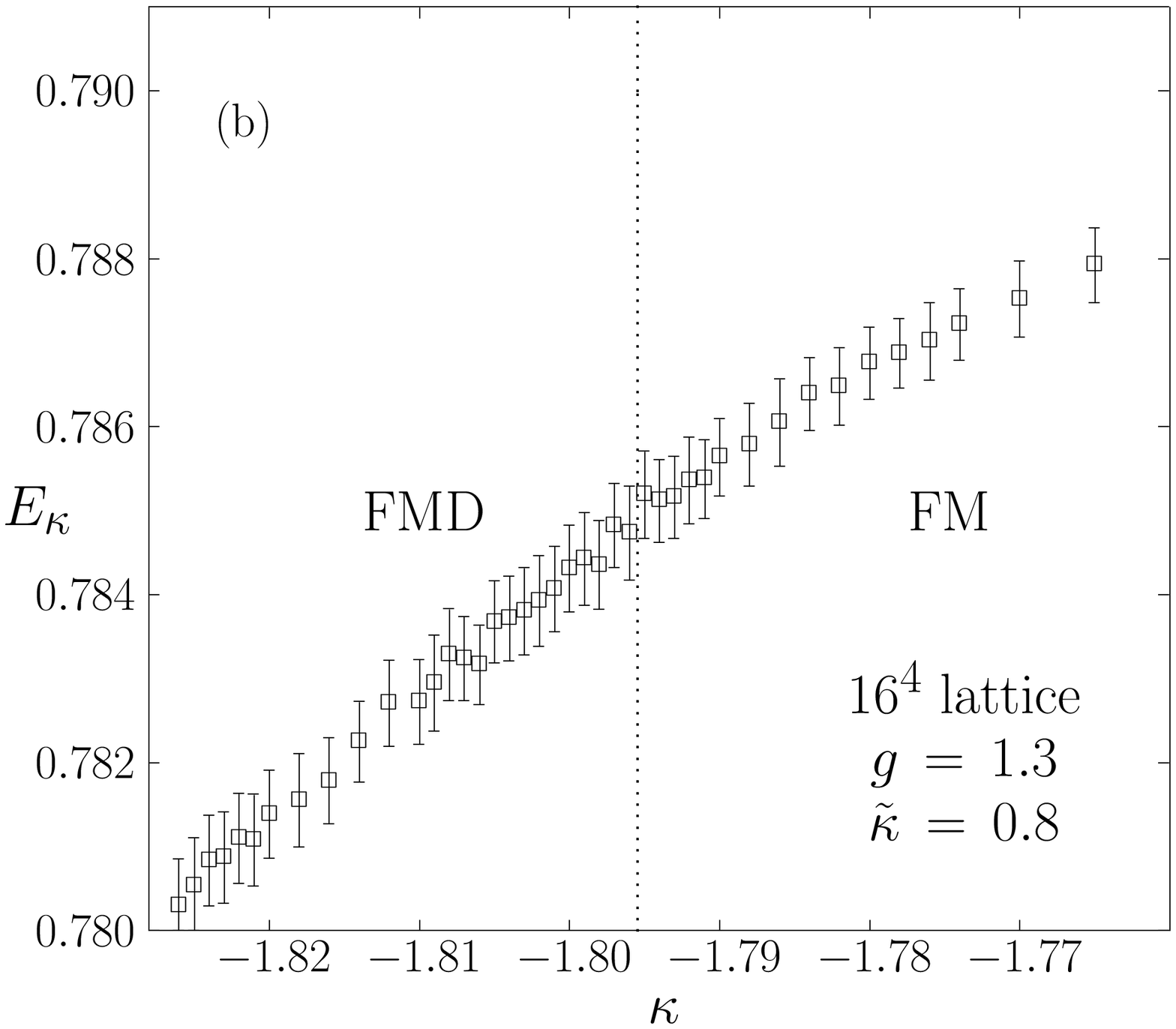,width=8.6cm,height=10.6cm}}
\vspace{-2.9cm}
\caption{ Continuous change in $E_{\kappa}$ (for large gauge
coupling) across FM-FMD transition (the dotted line) at larger 
$\tilde{\kappa}$, for (a) $10^4$ and (b) $16^4$ lattices} 
\label{ordr1}
\eef

Clearly there is a huge qualitative difference between figs.\ref{ordr}
and \ref{ordr1} in the nature of change of $E_{\kappa}$ versus $\kappa$
across the FM-FMD transition. Please note that the scales of the figs.
\ref{ordr}(a) and \ref{ordr1}(a), which show data for only the $10^4$
lattices, are exactly the same.  
In addition, although the figs. \ref{ordr} and \ref{ordr1} are shown only 
for $10^4$ and $16^4$ lattices, we have actually looked 
at all the observables on lattices from $4^4$ all the way  
upto $16^4$, and we have seen that the qualitative difference discussed
above gets only more pronounced at larger lattices. At
$\tilde{\kappa}=0.4$, as shown in figs. \ref{ordr}(a) and (b), the
discrete jump of $E_{\kappa}$ gets distinctly sharper on the $16^4$ lattice. 
On the other hand, at $\tilde{\kappa}=0.8$ as shown in fig. \ref{ordr1}(a), 
$E_{\kappa}$ is quite continuous across the FM-FMD transition on $10^4$ 
lattices. Even a fine resolution
of data points separated by
$\Delta\kappa=0.001$ does not show any discontinuity. As we go to $16^4$
lattices as shown in fig. \ref{ordr1}(b), we critically investigate only the 
region 
around the transition (which obviously shifts a little with the change of 
lattice size) and absolutely no discontinuity is found with our resolution.
In addition, a hint of a S-shape around the transition is visible which
promises to evolve into an inflexion with infinite slope at the transition
in the thermodynamic limit.

We have probed the gauge-fixed pure gauge system by quenched staggered
fermions with U(1) charge by measuring the chiral condensate
\be
\langle \bar{\chi} \chi \rangle_{m_0} = \frac{1}{L^4} \sum_x
\langle M^{-1}_{xx} \rangle \label{chcnd}
\ee
as a function of vanishing fermionic bare mass $m_0$. $M$ is the fermion 
matrix. The chiral condensates are computed
with the Gaussian noise estimator method \cite{gnem}. Anti-periodic
boundary condition in one direction is employed.

\bef
\vspace{-1.5cm}
\centerline{\psfig{figure=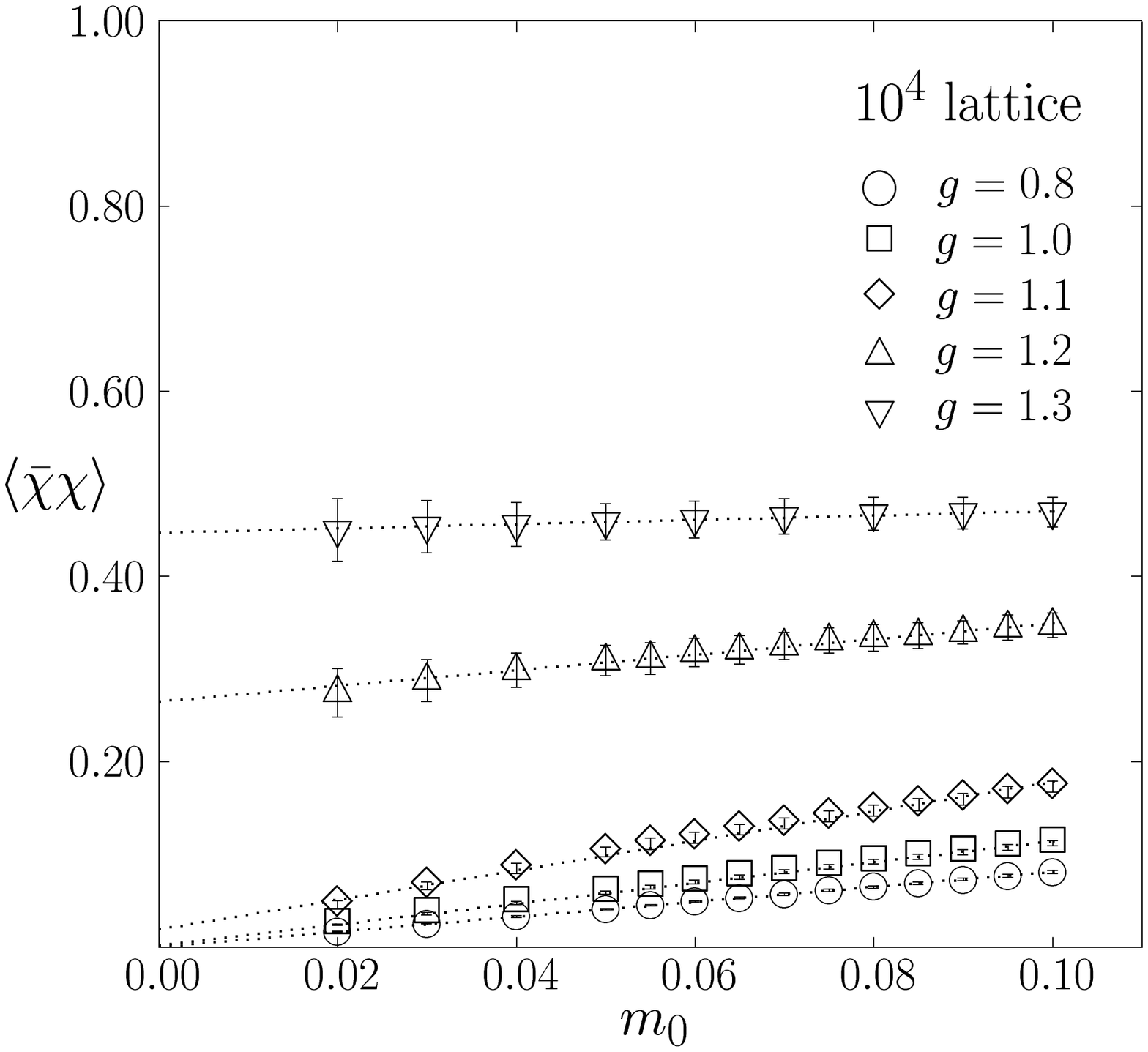,width=12.0cm,height=14.3cm}}
\vspace{-4.0cm}
\caption{{ Quenched chiral condensate on $10^4$ lattice as a
function of $m_0$ for different $g$ at $\tilde{\kappa}=0.4$.}} \label{chcds}
\vspace{-0.5cm}
\eef

\bef
\vspace{-2.5cm}
\centerline{\psfig{figure=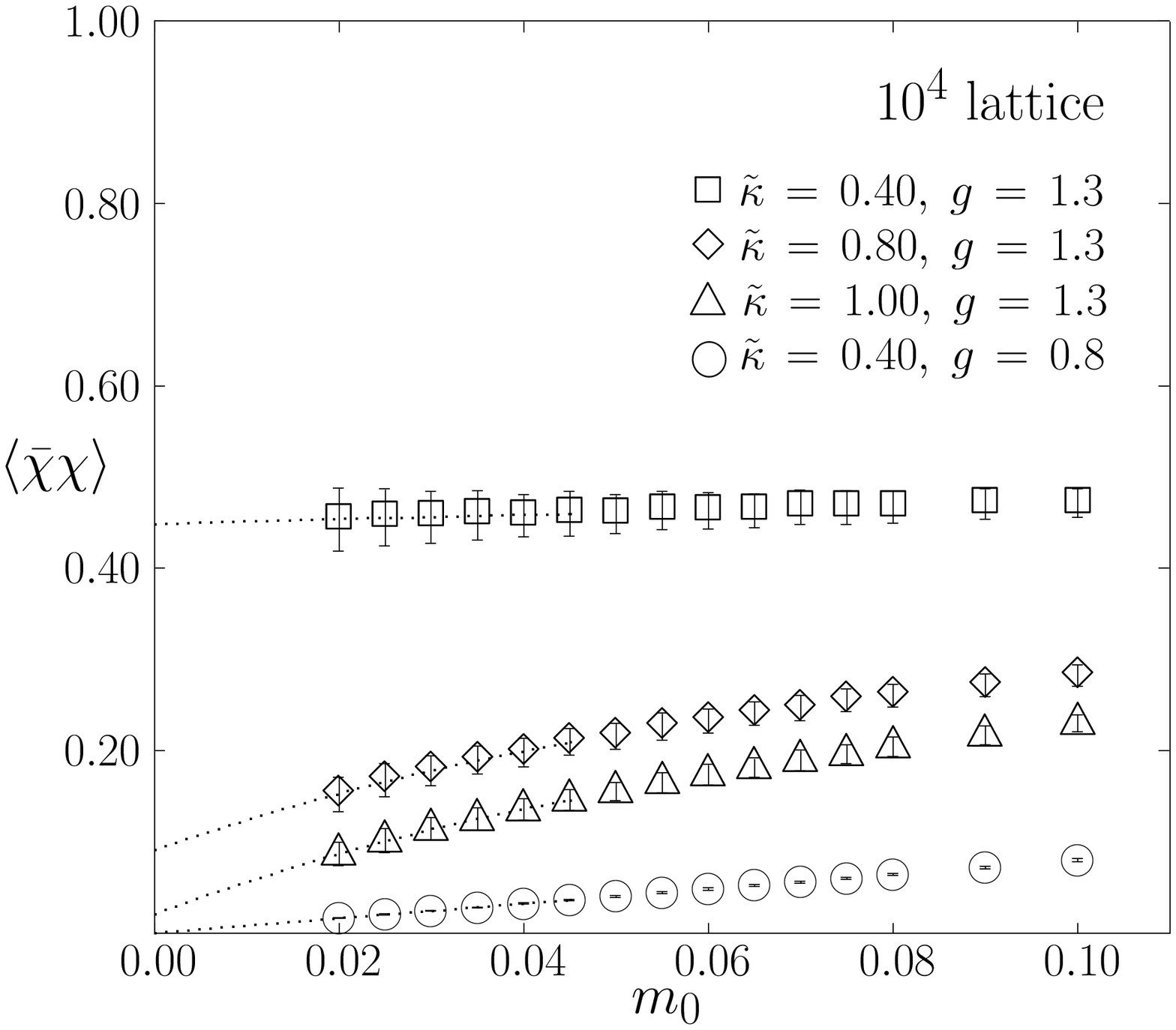,width=12.0cm,height=14.3cm}}
\vspace{-3.5cm}
\caption{{ Quenched chiral condensate on $10^4$ lattice as a
function of $m_0$ for different $\tilde{\kappa}$.}} \label{chcn0}
\vspace{-0.5cm}
\eef

Figure \ref{chcds} shows quenched chiral condensates near the
FM-FMD transition (remaining in the FM phase) obtained on $10^4$
lattice for different gauge couplings as a function of staggered
fermion bare mass $m_0$ at ${\tilde{\kappa}}=0.4$. Figure \ref{chcds}
clearly indicates that for weaker gauge couplings ($g < 1.1$)
the chiral condensates vanish in the chiral limit. For stronger gauge couplings  
($g > 1.1$) the chiral condensates are clearly not zero in the chiral 
limit. The dotted lines in fig. \ref{chcds} (and in fig. \ref{chcn0} to
follow) are only to guide the eye. It is to be noted that
at ${\tilde{\kappa}}\sim 0.4$ as $g$ changes from below 1.1 to above, the 
FM-FMD transition changes from continuous to 
first order (see fig.\ref{ordr}).
   
Figure \ref{chcn0} also shows chiral condensates near the FM-FMD transition 
again as function of the bare fermion
mass but this time for different values of $\tilde{\kappa}$. Here we
observe, interestingly, that for $\tilde{\kappa}\gtrsim 0.8$ and large gauge
coupling ($g=1.3$) the chiral condensates tend to vanish in the chiral limit.
It is to be noted again that at $g=1.3$ as $\tilde{\kappa}$ changes 
from roughly below 
0.8 to above, the FM-FMD transition changes from first order to
continuous. 

The above discussion strongly suggests that
inclusion of fermions here leads to a chiral phase transition that
intersects the FM-FMD phase transition. For a fixed
$\tilde{\kappa}$ there is chiral transition as $g$ is changed, and for a
fixed $g$ the chiral transition shows up as $\tilde{\kappa}$ is changed (the
third parameter $\kappa$ is used to stay on the FM-FMD transition). A
similar phenomenon occurs for the {\em order} of the FM-FMD transition with
respect to changes in $g$ and $\tilde{\kappa}$.
 
There is no chiral condensate in regions where we can take a continuum limit
in the pure gauge theory (which is the expected perturbative result). On the
other hand, chiral condensates appear where there is no continuum limit for
the pure gauge theory.

The tricritical point (or line in the 3 dimensional parameter space) where
the FM-FMD transition changes its order seems to be the likely place where the
chiral transition intersects the FM-FMD transition. The tricritical line is
therefore the only candidate where there is a possibility of a continuum
limit with nonperturbative properties like the chiral condendsate. 

\vspace{-0.3cm}

\section*{Conclusion}

With the particular regularization of compact pure U(1) gauge theory with an
extended parameter space, we have shown that there is clearly
a continuum limit for the whole range of the bare gauge coupling $g$.
Evidence of a continuum limit in other regularizations of a compact
lattice U(1) gauge theory is either absent, inconclusive
\cite{azco0,jiri0,flux} 
or dependent on inclusion of fermionic interactions \cite{azco1,kogut,chiu}.

Given the long history of speculation about a confining strong coupling
U(1) gauge theory and related issues of non-triviality, we have probed the
pure gauge system by quenched staggered fermions and found a clear evidence 
for a chiral phase transition. However, the region with a nonzero chiral
condensate does not allow a continuum limit. The continuum limit in the pure
gauge theory is only attained with no chiral condensate. This is consistent
with perturbative expectations.

We have found reasonable evidence to expect that the tricritical line at
which the the order of the FM-FMD phase transition changes 
in the pure gauge theory,
coincides with the line where the chiral phase transition intersects
the FM-FMD transition. This line is the only candidate for a possible
continuum limit with nonperturbative properties like chiral condensates.  
     
The authors thank M. Golterman and P.B. Pal for reading the manuscript and
comments. This work is funded by a project under the DAE. One of the authors
(SB) acknowledges the support of the U.S. Dept. of Energy under grant  
no. DE-FG02-93ER-40762.

\vspace{-0.3cm}


\end{document}